\documentclass[%
 aip,
 amsmath,amssymb,
 reprint,%
]{revtex4-1}

\usepackage{graphicx}
\usepackage{dcolumn}
\usepackage{bm}

\usepackage[utf8]{inputenc}
\usepackage[T1]{fontenc}
\usepackage{mathptmx}
\usepackage{xcolor}
\usepackage{physics}

\begin{document}

\preprint{AIP/123-QED}

\title{Interferometric speckle visibility spectroscopy (ISVS) for human cerebral blood flow monitoring }

\author{J. Xu} \thanks{These authors contributed equally to this work.}
\author{A. K. Jahromi} \thanks{These authors contributed equally to this work.}
\author{J. Brake} \thanks{These authors contributed equally to this work.}
\affiliation{Department of Electrical Engineering, California Institute of Technology}
\author{J. E. Robinson}%
\affiliation{Division of Biology and Biological Engineering, California Institute of Technology}%

\author{C. Yang}
\email{chyang@caltech.edu.}
\affiliation{Department of Electrical Engineering, California Institute of Technology}

\date{\today}

\begin{abstract}
    Infrared light scattering methods have been developed and employed to non-invasively monitor human cerebral blood flow (CBF). However, the number of reflected photons that interact with the brain is low when detecting blood flow in deep tissue. To tackle this photon-starved problem, we present and demonstrate the idea of interferometric speckle visibility spectroscopy (ISVS). In ISVS, an interferometric detection scheme is used to boost the weak signal light. The blood flow dynamics are inferred from the speckle statistics of a single frame speckle pattern. We experimentally demonstrated the improvement of measurement fidelity by introducing interferometric detection when the signal photon number is insufficient.We apply the ISVS system to monitor the human CBF in situations where the light intensity is $\sim$100-fold less than that in common diffuse correlation spectroscopy (DCS) implementations. Due to the large number of pixels ($\sim 2\times 10^5$) used to capture light in the ISVS system, we are able to collect a similar number of photons within one exposure time as in normal DCS implementations. Our system operates at a sampling rate of 100 Hz. At the exposure time of 2 ms, the average signal photon electron number is $\sim$0.95 count/pixel, yielding a single pixel interferometric measurement signal-to-noise ratio (SNR) of $\sim$0.97. The total $\sim 2\times 10^5$ pixels provide an expected overall SNR of 436. We successfully demonstrate that the ISVS system is able to monitor the human brain pulsatile blood flow, as well as the blood flow change when a human subject is doing a breath holding task.
\end{abstract}

\maketitle

\section{INTRODUCTION}

Over the last few decades, a variety of non-invasive optical schemes have been developed to study the cerebral blood flow (CBF) dynamics in human brains\cite{Jobsis77Science, Cheung01PMB, Huppert09AO, Eggebrecht14NatPhoton}, including near-infrared spectroscopy (NIRS) \cite{Jobsis77Science}, diffuse correlation spectroscopy (DCS)\cite{Cheung01PMB, Huppert09AO} and diffuse optical tomography (DOT)\cite{Eggebrecht14NatPhoton}. The 650–950 nm optical window has relatively low optical absorption and therefore enables light to penetrate through the skin, scalp, and skull and interact with the brain. Returning photons carry information about the CBF and can be used to infer the brain activity via neurovascular coupling\cite{Lou87AnnNeurol,  Dirnagl94AJPHCP}. In short, neurovascular coupling describes the connection between brain activity and CBF, since adequate CBF ensures sufficient oxygen is delivered to the tissue\cite{Buckley14Neurophot}. 

In recent years, there has been a renewed effort to use DCS for such measurements. DCS uses coherent red or near-infrared lasers as light sources and high-bandwidth single-photon counting modules (SPCM) as detectors. Compared to NIRS and DOT that measure reflected light intensity, DCS analyzes the dynamic scattering by monitoring intensity fluctuations. Therefore, DCS measures the flow dynamics rather than the hemoglobin concentration. Compared to the existing functional brain activity detection approaches, such as functional magnetic resonance imaging (fMRI) and electroencephalography (EEG), DCS offers an additional capability with some advantages. In terms of refresh rate, recent work from Wang \textit{et al.}\cite{Wang16BOE} showed a refresh rate of tens of Hz, which is able to well sample the pulsatile effect in CBF. Since the sampling bandwidth of fast fMRI is in the order of 1 Hz\cite{Lewis16PNAS} and common human cardiac signals (heartbeat) usually have a period of 1-3 Hz, the cardiac signals are not well sampled, which significantly contributes to the noise in fMRI\cite{Brooks13FHN}. Typical DCS systems are able to provide a spatial resolution of $\sim\!$ 1 cm\cite{Durduran04OL, Durduran14Neuroimage}. Non-invasive EEG, the gold standard technique for monitoring brain activity in terms of speed, provides signals from a mixture of multiple underlying brain sources. This results in low spatial resolution of 5–9 cm\cite{Burle15IJP}. Therefore, DCS lies in the niche between fMRI and EEG, where fMRI has high spatial resolution ($\sim\!$ 2 mm)\cite{Durduran14Neuroimage} but relatively low temporal resolution ($\sim\!$ 1 Hz) and EEG has high temporal resolution ($>$\! 1 kHz)\cite{Zhang14CognNeurodyn} but low spatial resolution (5–9 cm).

The performance of DCS is ultimately constrained by the total collected photon budget required for a reasonable signal-to-noise ratio (SNR). Since DCS exploits only a few speckles grains from the scattered light, in order to get a sufficient number of signal photons for a relatively accurate measure on the blood flow dynamics, the required measurement time of the detector for one data point (one measurement of CBF) is typically no less than tens of milliseconds. There is therefore a tradeoff between the measurement time and the sensitivity of the system: a high SNR measurement requires relatively long measurement time, which results in a relatively low sampling rate. This tradeoff is fundamentally caused by the limitation of photon budgets and can be mitigated by using multi-channel DCS, which in turn scales up the costs of the system\cite{Zhou18Optica, Wang16BOE}. Furthermore, the multi-channel DCS also scales up the requirement of data throughputs, which also brings technical issues in practical implementations.   

To tackle the limitation of the shortage of signal photons, people use a camera sensor with thousands to millions pixels as the detector in place of the SPCM. By using a camera sensor, the system can collect more photons with the same measurement time compared to DCS, or collect the same number of photons with a shorter measurement time. Because the  number of pixels is large while the data throughput is limited, the temporal sampling speed is typically not sufficient to well sample the temporal intensity fluctuation. Even though, the temporal dynamics of blood flows can be inferred from the speckle statistics of the captured frame. This method is termed speckle visibility spectroscopy (SVS), or speckle contrast spectroscopy\cite{Dunn01JCBFM, Bandyopadhyay05RSI, Dunn12ABE, Valdes14BOE, Zhao20MM}. In the SVS-based approach, a whole speckle pattern containing many speckle grains is captured using a detector array (e.g., a CCD camera), and the statistics of the blurred speckle pattern is used to calculate the speckle decorrelation time. With a given camera integration time that is longer than the speckle decorrelation time, different speckle decorrelation times result in speckle frames with different extent of blurring. Since a speckle pattern recorded by a detector array can contain thousands to millions of speckle grains, the photon budget limitation is mitigated. In this case, a single decorrelation time measurement does not require tens of millisecond as compared to DCS, therefore resulting in a higher refresh rate. 

While SVS relaxes the requirements for temporal fidelity, it suffers from camera noise as commercial camera sensors are usually noisier than SPCMs when the signal light intensity is low. When detecting “deep photons” — those interacting with tissues at larger depths ($>\!$ 1 cm) — the amount of reflected photons reaching the detector array can be less than 1 photon/pixel within the camera exposure time. In these cases, the camera noise would overwhelm the detected deep photons. The direct way to overcome this problem would be to increase the camera exposure time. Similar to the aforementioned tradeoff between SNR and measurement time, the increase of exposure time will result in a decrease in the refresh rate. 

Here we propose and demonstrate the idea of interferometric SVS, or ISVS, which circumvents the camera noise problem and is able to measure the blood flow dynamics even when the number of available signal photons is limited (below 1 photon electron per pixel). Interferometric detection is able to overcome camera noise by boosting the weak signal term in the heterodyne cross-term, and as such, the ISVS system is able to achieve a reasonable SNR even when the mean pixel value number from the signal light is smaller than 1. A novel interference speckle spatial statistics analysis is derived to quantify the speckle decorrelation time. Another advantage brought by ISVS is that the interferometric measurement provides the information of complex field decorrelation time, therefore the Siegert relation\cite{Siegert1943} as well as the corresponding Gaussian statistics assumptions required in typical DCS calculations, are not required in ISVS. This makes ISVS valid even when the conventional speckle statistical property is broken due to static scattering or unexpected background light received by the camera sensor.  

By using ISVS, we demonstrate high speed (100 Hz) non-invasive \textit{in-vivo} cerebral blood flow monitoring on the human forehead, under the condition where the number of signal photon electrons per pixel is $\sim\!$ 0.95. In this case, the camera sensor noise is $\sim\!$1.2 photon electrons per pixel. The direct measurement will yield an SNR of $\sim\!$0.61 for each pixel due to the photon electron shot noise and the detector noise. The overall SNR with $\sim 2\times 10^5$ pixels is an expected value of $\sqrt{0.61 \times 2\times 10^5} \approx 349$. By using ISVS, the detector noise effect can be mitigated and the single pixel measurement SNR is dominant by shot noise. Therefore, in the interferometric detection regime, the single pixel SNR should be $\sqrt{0.95}\approx 0.97$ due to shot noise, and the use of $\sim 2\times 10^5$ camera pixels provides an expected SNR of 436. While the SNR improvement by using interferometric detection is modest for this camera, this SNR improvement is, nevertheless, desirable. We expect this ability of interferometric detection to provide shot-noise limited detection SNR would be even more desirable in future ISVS experiments where near-infrared cameras with noisier characteristics are used in tandem with longer wavelengths.   We also design a breath holding task for a human subject and implement ISVS to monitor the CBF, showing that the relative CBF (rCBF) changes in accordance to brain stimulation caused by the task can be revealed by ISVS.

\section{RESULTS}
\subsection{Principle}

\begin{figure}
\centering\includegraphics[width=8.6cm]{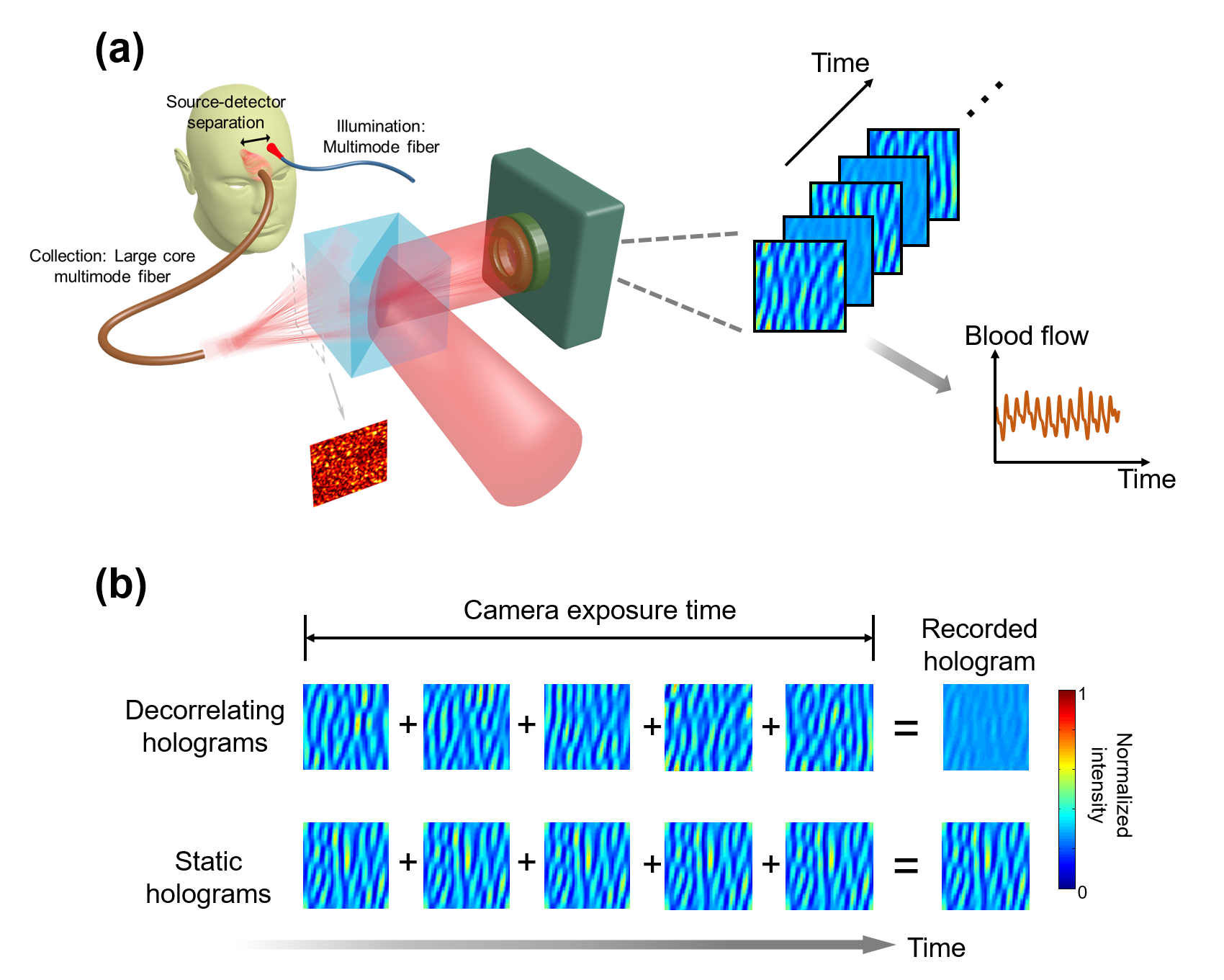}
\caption{\label{Figure 1}Principle of ISVS. (a) Schematic of ISVS setup. A beam illuminates the subject and the diffused light is collected by the optical system interferes with a reference beam on a camera. The sequence of the interference patterns is used to reconstruct the signal trace that contains blood flow information. (b) Difference between decorrelating and static diffused optical field recorded in a single camera frame. When the diffused optical field is decorrelating, the camera integration sums the independent holograms and results in a low fringe visibility hologram; when the diffused optical field is static, the same hologram lasts during the whole camera integration time and results in a high fringe visibility hologram.}
\end{figure}

The ISVS principle is based on an interferometer, where the weak diffused light containing the information is boosted by a reference beam and recorded by the camera (See Fi. \ref{Figure 1}(a)). The image sequence recorded on the camera can be reconstructed to provide a sequence that contains blood flow information, where each camera frame corresponds to one data point in the reconstructed sequence. Assuming the diffused light has a limited spatial bandwidth, the complex field of the diffused light is obtained via an off-axis holography configuration\cite{Takeda82JOSA}
, by directly reconstructing the measured output intensity of the interferometer. In this configuration, the total instantaneous interference pattern It at the position $\mathbf{r}\!=\!(x,y)$ in the observation plane is
\begin{eqnarray}\label{Eq1}
I_\mathrm{t}(\mathbf{r})&&=|E_\mathrm{R}|^2+|E_\mathrm{S}(\mathbf{r})|^2+2|E_\mathrm{R}||E_\mathrm{S}(\mathbf{r})|\cos{\Big(k_0x\sin{\theta}+\phi_\mathrm{S}(\mathbf{r})\Big)}\nonumber\\&&
=I_\mathrm{R}+I_\mathrm{S}(\mathbf{r})+2\sqrt{I_\mathrm{R}I_\mathrm{S}(\mathbf{r})}\cos{\Big(k_0x\sin{\theta}+\phi_\mathrm{S}(\mathbf{r})\Big)},
\end{eqnarray}
where $E_\mathrm{R}$ is the complex field of the plane wave reference beam, $E_\mathrm{S}(\mathbf{r})$ is the complex field of the diffused light, $\phi_\mathrm{S}(\mathbf{r})$ is the phase difference between the reference beam and the diffused light field, $k_0$ is the wave-vector in free space, and $\theta$ is the tilt angle of the oblique reference beam. Thanks to the off-axis holography, the Fourier transform of $I_\mathrm{t}(\mathbf{r})$ in Eq.\! (\ref{Eq1}) provides three separate lobes (see Supplementary Fig. 1A), where the central lobe is the Fourier transform of the DC terms (first two terms in Eq.\! (\ref{Eq1})) and the two side lobes are the Fourier transform of the interference term (third term in Eq.\!\!\! (\ref{Eq1})). The reconstructed signal $I_\mathrm{rec}(\mathbf{r})$ is with $I_\mathrm{R}\!=\!|E_\mathrm{R}|^2$ and $I_\mathrm{S}(\mathbf{r})\!=\!|E_\mathrm{R}(\mathbf{r})|^2$. Here $I_\mathrm{rec}(\mathbf{r})$ contains the information of the diffused light. It is worth noting that the instantaneous analysis is valid for a given time (e.g. $t\!=\!0$), whereas the detector has a non-zero integration time to record the intensities. Nevertheless, this analysis is indicative of the SNR advantage through interferometry: for most speckle grains, the intensity of a speckle grain is smaller than the reference beam intensity, resulting in $I_\mathrm{rec}(\mathbf{r})\!>\!I_\mathrm{S}(\mathbf{r})$. Thus the signal experiences a significant boost, which can salvage extremely weak signals from being buried under the detector noise. Note that this basic argument is not accurate for the intensities measured in the experiment, since the detector measures the intensities for a given integration time, rather than giving an instantaneous value. A careful analysis below shows the exposure time T reduces the energy of the heterodyne terms by a less-than-unity ‘visibility factor’. 

We now move forward to quantitatively demonstrate how ISVS is able to distinguish between decorrelation events of different timescales. The analysis is dependent on interference speckle spatial statistics, which is exploited for the first time in CBF measurements. In Fig.\! \ref{Figure 1}, the interference pattern recorded by the camera is
\begin{eqnarray}\label{Eq2}
I(\mathbf{r})~&&=\int_0^T \Big[|E_\mathrm{R}|^2+|E_\mathrm{S}(\mathbf{r},t)|^2+2|E_\mathrm{R}||E_\mathrm{S}(\mathbf{r},t)|\nonumber\\&&\times\cos{\Big(k_0x\sin{\theta}+\phi_\mathrm{S}(\mathbf{r},t)\Big)}\Big]dt,
\end{eqnarray}
where $t\!=\!0$ defines the beginning of the exposure and T is the exposure time. Since the third interference term in the integral has a different location on Fourier space, one can filter it out to get the complex field information. Let us define the interference signal as $H(\mathbf{r})\!=\!\frac{1}{T}\int_0^T 2|E_\mathrm{R}||E_\mathrm{S}(\mathbf{r},t)|\nonumber\cos{\!\phi_\mathrm{S}(\mathbf{r},t)}~dt$. It can be rewritten as $H(\mathbf{r})\!=\!\frac{1}{T}\int_0^T |E_\mathrm{R}||E_\mathrm{S}(\mathbf{r},t)|\nonumber e^{i\phi_\mathrm{S}(\mathbf{r},t)}dt+\!\frac{1}{T}\int_0^T |E_\mathrm{R}||E_\mathrm{S}(\mathbf{r},t)|\nonumber e^{-i\phi_\mathrm{S}(\mathbf{r},t)}dt$. We can then pick the first one of the phase conjugated pairs and define the ISVS signal $S(\mathbf{r})$ as 
\begin{equation}\label{Eq3}
S(\mathbf{r})\!=\!\frac{1}{T}\int_0^T |E_\mathrm{R}||E_\mathrm{S}(\mathbf{r},t)|\exp\Big(i\phi_\mathrm{S}(\mathbf{r},t)\Big)~dt.
\end{equation}
The second moment of $S(\mathbf{r})$ will contain the field decorrelation function $g_1(t)\!=\!\langle E_\mathrm{S}(\mathbf{r},t_0)~\! E^*_\mathrm{S}(\mathbf{r},t_0\!+\!t)\rangle_{t_0}/\langle|E_\mathrm{S}(\mathbf{r},t_0)|^2\rangle_{t_0}$ after the mathematical derivation shown in Eq. (\ref{Eq4}) below (See more details in Supplementary Note D and Ref. \citenum {Hussain18JB}):
\begin{widetext}
\begin{equation}\label{Eq4}
\langle|S(\mathbf{r})|^2\rangle\!=\!\frac{I_\mathrm{R}}{T^2}\bigg\langle\int_0^T \int_0^T dt_1 dt_2~E_\mathrm{S}(\mathbf{r},t_1)\exp\Big(\!i\phi_\mathrm{S}(\mathbf{r},t_1)\!\Big)E_\mathrm{S}(\mathbf{r},t_2)\exp\Big(\!\!-\!i\phi_\mathrm{S}(\mathbf{r},t_2)\!\Big) \bigg\rangle\!=\!\frac{I_\mathrm{R}\bar{ I_\mathrm{S}}}{T}\int_0^T 2(1-\frac{t}{T})g_1(t)~dt,
\end{equation}
\end{widetext}
where $\langle\bullet\rangle_{t_0}$ denotes the expected value over $t_0$, $\langle\bullet\rangle$ denotes the expected value over space, $I_\mathrm{R}$ is the intensity of the reference beam, and $\bar{I_\mathrm{S}}$ is the mean intensity of the signal beam. Both of $I_\mathrm{R}$ and $\bar{I_\mathrm{S}}$ can be calibrated before the experiment. We then define the visibility factor
\begin{equation}\label{Eq5}
F\!=\!\frac{\langle|S(\mathbf{r})|^2\rangle}{I_\mathrm{R}\bar{ I_\mathrm{S}}}\!=\!\frac{1}{T}\int_0^T 2(1-\frac{t}{T})g_1(t)~dt
\end{equation}
which ranges from 0 to 1 for different $g_1(t)$.

As we can see from Eq. (\ref{Eq5}), the visibility factor $F$ is a function of $g_1(t)$ and the camera exposure time $T$. If the sample is static (i.e., $g_1(t)\!=\!1$ for $0<t<T$), $F\!=\!1$, indicating that the interference fringes have high contrast. If the complex field of the diffuse light decorrelates very quickly compared to $T$, (i.e.,  $g_1(t)\!=\!1$ for $0<t<\tau$ and $g_1(t)\!=\!0$ for $\tau<t<T$ where $\tau\ll T$), $F\!=\!\frac{2\tau}{T}$, indicating that the interference fringes have low contrast. This matches with the intuition that multiple decorrelation events happening within the camera exposure time blur out the interference fringes and yield a low fringe visibility.

Fig.\! \ref{Figure 1}(b) intuitively shows the relation between the visibility factor and the decorrelation time scale. Let us consider two cases: a static optical field (top row in Fig.\! \ref{Figure 1}(b) and a decorrelating optical field (bottom row in Fig.\! \ref{Figure 1}(b). If the optical field is static, the amplitude and phase of one speckle grain is fixed within the camera exposure time $T$ and thus the interference fringes have high visibility. If the optical field is decorrelating within the camera exposure time, the pattern recorded on the camera will be the summation of different interference patterns with difference amplitudes and phases. The summation of the patterns from uncorrelated sample field will cause a low fringe visibility. As the decorrelation time shrinks relative to the camera exposure time $T$, the number of decorrelating events within $T$ grows, therefore leading to lower fringe visibility.

\subsection{System characterization and measurement fidelity improvements}

To verify and characterize the performance of the system, we used two ground glass diffusers (Thorlabs, DG20 Series) as the samples. The schematic is shown in Fig.\! \ref{Figure 2}(a) (more details in Supplementary Fig. 2A). The first one had controlled rotating speeds and the other one is static. A laser beam illuminated the rotating diffuser and a range of rotation speeds were applied. The scattered light from the first rotating diffuser illuminated the second static diffuser and was collected by the optical system. The static diffuser was used to eliminate the speckle pattern “smearing” effect that is present when using a single rotating diffuser. The decorrelation times were computed by measuring the time traces of the intensity fluctuations using a single photon counting module (SPCM) in the optical setup. Then, by mapping rotation speed to the measured decorrelation time, we were able to simulate a dynamic sample with configurable decorrelation times. 

First, we verified the mathematical model of the ISVS visibility factor by experimental results. The blue curve in Fig.\! \ref{Figure 2}(b) depicts the measured ISVS visibility factors under different camera exposure times ($T$) to decorrelation time ($\tau$) ratios, $T/\tau$, where the decorrelation time $\tau$ of the scattering light was measured by analyzing the autocorrelation of the time traces captured by the SPCM in parallel to the ISVS measurement. The error bars on vertical axis were calculated from 10 ISVS measurements, and the error bars on horizontal axis were calculated from 10 DCS measurements. The red curve is the theoretical model calculation (See Supplementary Eq. 5), showing good correspondence with the experimental results. 

Using the theoretical model, we then mapped the measured visibility factors to decorrelation times (Fig.\! \ref{Figure 2}(c) vertical axis) and compared them to the DCS measured decorrelation times (Fig.\! \ref{Figure 2}(c) horizontal axis). The black dotted curve is the line with the slope equal to 1. The agreement between the ISVS and DCS measured speckle decorrelation times shows the ability of the system to quantitatively measure decorrelation time.

Finally, we showed that the signal boost from the interferometric detection scheme resulted in a measurement with higher fidelity compared to direct detection, when the number of signal photons from the sample was low. In this experiment, we performed direct SVS measurements and used the model from Ref. \citenum{Bandyopadhyay05RSI} to calculate the scattered light decorrelation time. As shown in (Fig.\! \ref{Figure 2}(d), when the number of photon-electrons excited by the signal beam was large ($>5$ e\textsuperscript{-}/pixel), both SVS and ISVS provided results close to the one in the DCS measurement, which served as the ground truth. As the signal beam became weaker, SVS measured decorrelation times deviated from the ground truth and got larger. In the low photon budget regime (e.g., $<1$ e\textsuperscript{-}/pixel) where mean intensity on the camera was low, the detector noise dominated the spatial speckle fluctuation, resulting in a high contrast, i.e., the ratio between the standard deviation and the mean value of the frame is high. For SVS, this camera-noise caused high contrast pattern may have a similar contrast as a frame with high contrast speckles. In this case, the model will inaccurately provide a long decorrelation time that is close to a static speckle pattern. In comparison, the interference term of ISVS is not significantly impacted by the camera noise and ISVS should provide an accurate measurement even when the detected signal photon count is low.

\subsection{System operation on rats}

We implemented an ISVS system to monitor blood flow in rats. The monitoring was performed in the dorsal skin and brain in a reflection-mode configuration. A nude rat was first anesthetized and then its dorsal skin was shaved and mounted to a clip device. A collimated laser beam illuminated the skin and the light diffused by the skin was detected by the ISVS system. The pulsatile effect of the blood flow from the dorsal skin flap measured by the ISVS system is presented in Supplementary Fig. 3A. Additionally, another nude rat was anesthetized and a portion of its scalp removed. A collimated laser beam illuminated on the skull and the light diffused from the skull and brain was detected by the ISVS system. The pulsatile CBF results from ISVS measurement are shown in Supplementary Fig. 3B. In the rat brain experiment, the bregma area was imaged by the system (Supplementary Fig. 3C) and the central part of the imaging area was $\sim\!$ 1 cm from the laser illumination spot. In both dorsal skin and brain blood flow measurements, a customized pulse monitor was synchronized with the optical system and monitored the blood flow in the tail, confirming that the detected pulsatile ISVS signals indeed originate from the pulsatile nature of the heart beating. A schematic of the experimental setup is shown in Supplementary Fig. 2B and details of the anesthesia and surgery are in Supplementary Note B. All of the procedures and the dosage of chemicals followed the protocols of the Institutional Animal Care and Use Committee at California Institute of Technology. Animal husbandry and all experimental procedures involving animal subjects were approved by the Institutional Animal Care and Use Committee (IACUC) and by the Office of Laboratory Animal Resources at the California Institute of Technology under IACUC protocol 1770-18.

\begin{figure}
\centering\includegraphics[width=8.6cm]{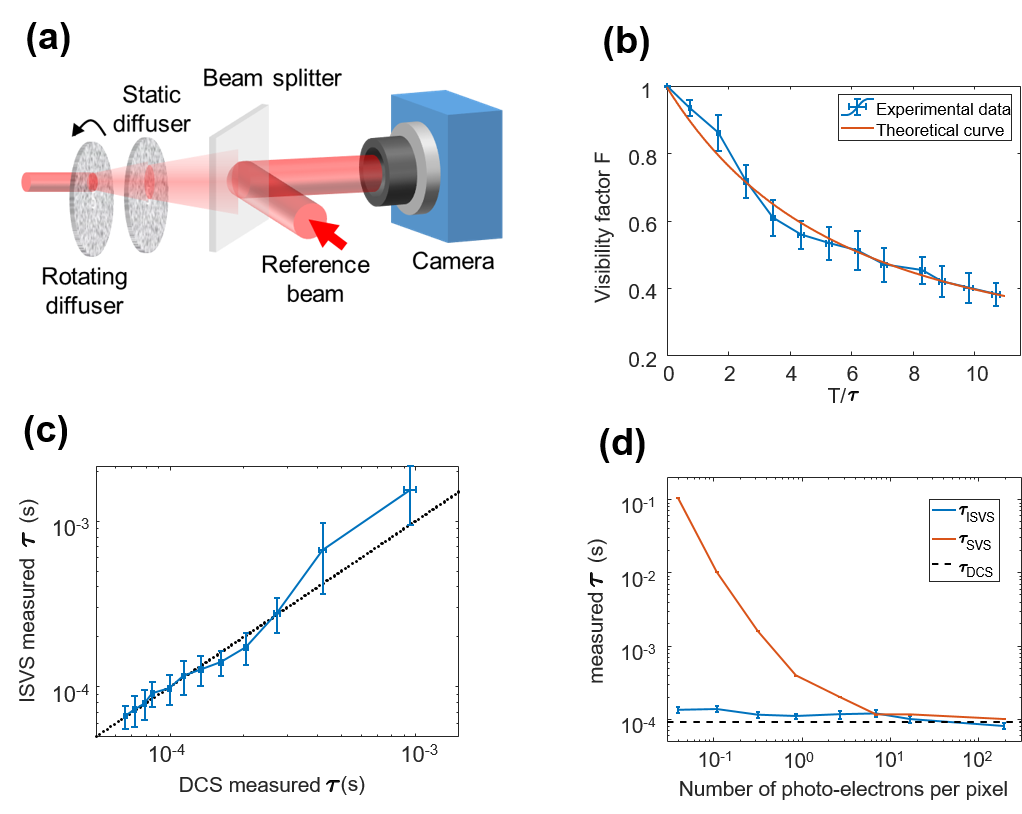}
\caption{\label{Figure 2}Characterization of ISVS by using rotating diffusers. (a) Schematic of the characterization experiment using two rotating diffusers. (b) Experimentally measured visibility factor F in different $T/\tau$ (blue curve), and the theoretical model curve (red curve). The vertical error bars are calculated from 30 ISVS measurements, and the horizontal error bars are calculated from 10 DCS measurements. (c) The comparison between the ISVS measured $\tau$ and DCS measured $\tau$. The blue curve is the experimental result, and the black dotted line is the line of unity slope. (d) Comparing measured $\tau$ from interferometric (ISVS, blue curve) and direct (SVS, red curve) detection in different light intensities. The horizontal axis shows the mean number of photo-electrons on each pixel of the camera. The error bars are calculated from 30 ISVS and SVS measurements. The decorrelation time measured by DCS (black dashed line) serves as the ground truth.}
\end{figure}

\subsection{Human experiment design and operation}

To demonstrate the performance of ISVS for \textit{in vivo} measurements of CBF dynamics, we implemented an ISVS system on human subjects, as depicted in Fig.\! \ref{Figure 3}(a). The schematic of a more detailed optical setup can be found in Supplementary Note A and Supplementary Fig. 2C. A beam splitter was added the split the light onto a single-photon-counting-module (SPCM) as a reference for photon count rates, as well as a DCS measurement. A healthy adult human subject sat on a medical chair with his head placed inside a head immobilizer. Non-contact source and detector fibers were mounted above the subject forehead over the prefrontal cortex area. To have a sufficient number of illumination photons for this measurement, the laser beam from the 671-nm source was coupled to a multimode fiber (Thorlabs M31L02, $\sim\!$ 3000 modes). The collimated 56 mW laser beam with a 6-mm spot size results in a $<2 \mathrm{mW/mm^2}$ irradiance for skin exposure – within the limit stipulated by American National Standard Institute (ANSI). The diffused light at various source-detector (S-D) separations (S-D = 1.5 cm and 0.75 cm) was then collected via a large core multimode fiber (Thorlabs M107L02, core diameter 1.5 mm) containing $\sim\!$ 6 million modes. The output of this fiber was channeled through the sample arm of the interferometer. About 2$\times10^5$ camera pixels are used to capture the photons. A human protocol comprising of all detailed experimental procedures were reviewed and approved by the Caltech Institutional Review Board (IRB) under IRB protocol 19-0941, informed consent was obtained in all cases, and safety precautions were implemented to avoid accidental eye exposure.

We first demonstrated that the ISVS system was able to monitor the blood flow in humans when the reflected light signal was low. When the S-D separation was 1.5 cm, the photon count rate read by the SPCM was $\sim\!$ 1500 counts/second, while the dark count rate of the SPCM was $\sim\!$ 180 counts/second. This photon rate is  $\sim\!$ 100 fold less than the light budget in conventional DCS experiments\cite{Wang16BOE}. Due to the experimental resources, the laser wavelength is 671 nm, at which biological tissue has a higher absorption coefficient than wavelength 785 nm in common DCS settings. The intensity decorrelation curve $g_2(t)$ is shown in Fig.\! \ref{Figure 3}(b) after a DCS measurement by the SPCM with a measurement time of 50 \!s. To obtain the pulsatile signal from the DCS measurement, the SPCM trace is divided into sub traces with the time duration of 10 ms for each sub trace. The blood flow index (BFI) is calculated from each sub trace, based on the tissue scattering parameters used in Ref. \citenum{Durduran10RPP}. As shown in Fig.\! \ref{Figure 3}(c), there is no obvious pulsatile signal retrieved from the DCS measurement. Therefore, under this experimental condition, the photon count rate was not sufficient for DCS to monitor the pulsatile signal with a reasonable SNR. The ISVS system with a camera exposure time of 2 ms and an FPS of 100 Hz yielded a pulsatile signal trace, shown by the blue color line in Fig.\! \ref{Figure 3}(d). The filtered signal trace is the red color line in Fig.\! \ref{Figure 3}(d). The measured ISVS visibility factor was used to calculate the decorrelation time, and the decorrelation time was used to calculate the BFI. The raw and filtered BFI traces are presented in blue and red curves in Fig.\! \ref{Figure 3}(e), respectively. The Fourier transform of the raw BFI trace is shown in Fig.\! \ref{Figure 3}(f), and the heart-beat frequency $\sim\!$ 1.1 Hz and its harmonics are highlighted. In this experimental configuration, the average photon electron number of the signal beam on each camera pixel was $\sim\!$ 0.95, where the detector noise was $\sim\!$ 1.2 photon electrons. In this case, the SNR for each pixel was $\sim\! 0.95/\sqrt{0.95+1.2^2} = 0.61$  with a direct measurement. With a shot noise dominant interferometric measurement, the single pixel SNR can achieve 0.97. By using $\sim 2\times 10^5$ camera pixels, the overall SNR can be scaled up by $\sqrt{2\times 10^5 }\approx$ 447 times to 436. It is worth mentioning that, due to the dynamic scattering interaction between the light and the blood flow, the contrast of the speckle pattern was much lower than unity, predicted by the static speckle statistics. This in fact makes direct detection even harder to retrieve the contrast of the pattern and the information of the blood flow.  
In order to demonstrate the capability of ISVS for human brain activity detection, we established a straightforward brain stimulation experiment (breath-holding task) for a healthy human subject. The experiment was based upon the fact that the stoppage of the subject’s breathing increases the CBF due to vasomotor reactivity from rising pCO\textsubscript{2}\cite{Harper65ANS, Hesser68RP}. CO\textsubscript{2} pressure will raise during breath holding\cite{Hesser68RP} due to metabolism, and the rising CO\textsubscript{2} pressure can cause the increase of blood flow\cite{Harper65ANS}. In the experiment, the subject first breathed normally for $\sim\!$ 2 minutes, then did an exhalation breath holding for $\sim\!$ 15 s, and finally started to recover by having normal breathing again.

The S-D separations in the breath holding task were set to be 1.5 cm and 0.75 cm. Since the $\sim\!$ 1 Hz pulsatile blood flow change naturally presented in the measurement, we set the sampling rate as 12 Hz. At this sampling rate the pulsatile signal was present, where data processing could filter it. The system recorded the $\sim\!$ 10 s before the breath holding, the entire $\sim\!$ 15 s of breath holding and the $\sim\!$ 16 s after the breath holding. We performed ISVS measurements in 3 cases: (i) An S-D separation of $d_1\!=\!1.5$ cm while the subject went through a breath-holding task, (ii) an S-D separation of $d_1\!=\!0.75$ cm while the subject went through a breath-holding task, and (iii) an S-D separation of $d_2\!=\!1.5$ cm while the subject breathed normally. The representative recorded traces for case (i), (ii) and (iii) are shown in Fig.\! \ref{Figure 3}(g) and Fig.\! \ref{Figure 3}(h), respectively. For each case, five repetitive experiments were conducted to avoid single experiment outlier. 

\begin{figure*}
\centering\includegraphics[width=14.6cm]{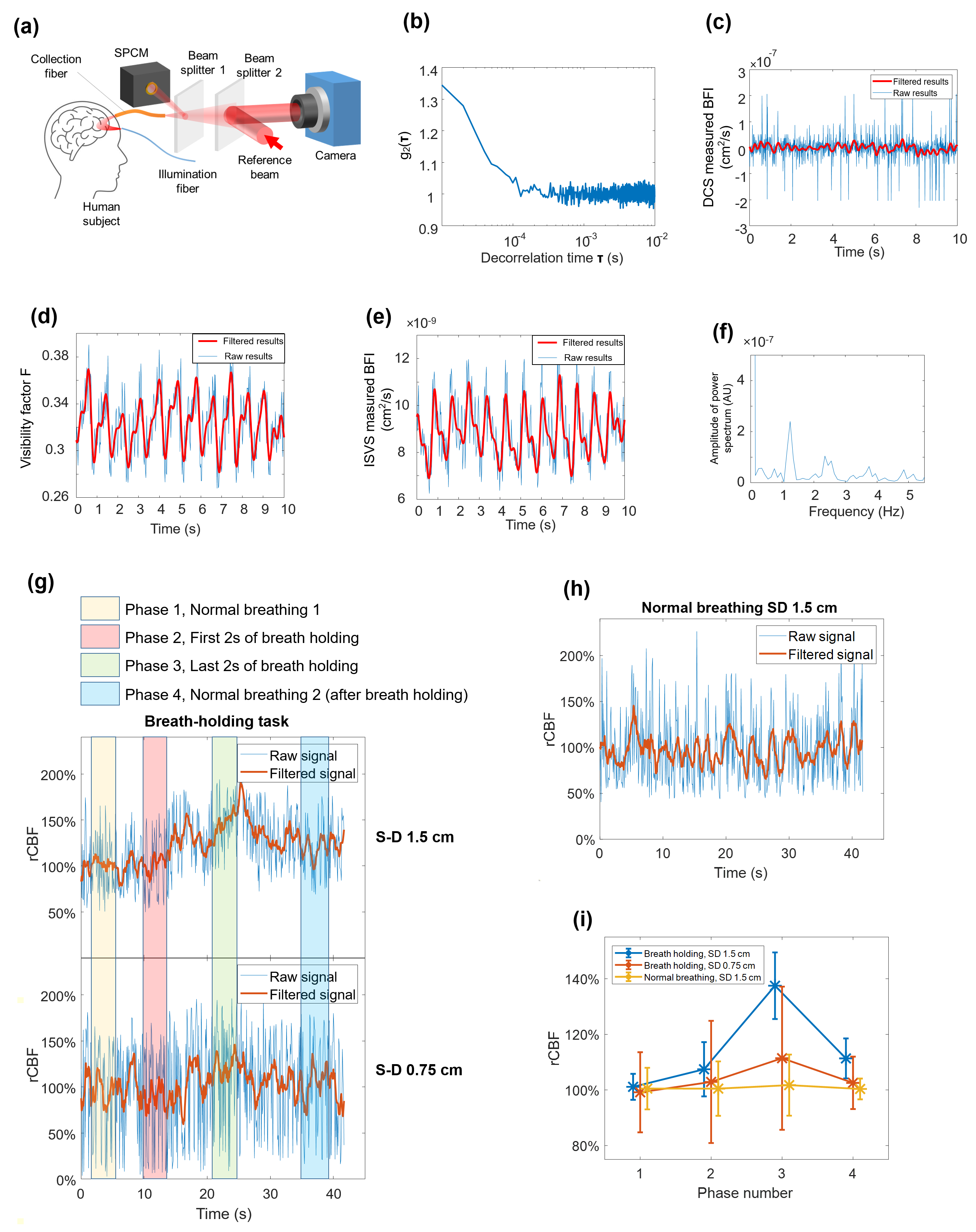}
\caption{\label{Figure 3}ISVS experimental results on a human subject. (a) Schematic of the human experiment. The illumination came from a multimode fiber, and the diffused light was collected by a large core multimode fiber. The output light of the large core multimode fiber was directed into the ISVS setup, where the diffused light and the reference beam were combined by a beam splitter and recorded by the camera. (b) The intensity correlation function $g_2(t)$ from a SPCM recording trace with time duration of 50 s. The decorrelation time is $\sim\!$ 50 $\mathrm{\mu s}$. (c) The BFI trace calculated from DCS. The entire intensity fluctuation trace is divided into sub traces, then the BFI is calculated from each sub trace. The BFI sampling rate is 100 Hz. (d) The ISVS visibility factor measured at 100 Hz on the forehead of the human subject. The blue curve shows the original data points, and the red curve is a low-pass filtered version of the blue curve. (e) Blood flow index calculated from the visibility factor. The blue curve was calculated from the original data points, and the red curve was low-pass filtered from the blue curve. (f) A Fourier transform of the blue curve in (e) shows the heart beat frequency at $\sim\!$ 1.1 Hz and its harmonics. (g) ISVS measured rCBF traces when the human subject was doing the breath holding task. The subject exhaled and started holding their breath at $\sim\!$ 10 s and kept it until around $\sim\!$ 15 s, after which the normal breathing was resumed. (h) ISVS measured rCBF trace when the human subject breathed normally. (i) Statistical analysis of rCBF change at different phases. Each data point and the error bar were calculated from 5 measurement curves from one human subject.}
\end{figure*}

To see the rCBF change due to the breath holding task, the mean values and standard deviations of rCBF during 2–4 s (Phase 1), 10–12 s (Phase 2, first several seconds of breath holding), 22–24 s (Phase 3, last several seconds of breath holding) and 37–39 s (Phase 4) were calculated and plotted, as shown in Fig.\! \ref{Figure 3}(i). The increase of rCBF values in Phase 3 at the S-D separation of 15 mm (case (i)) is clearly shown by the blue line in Fig.\! \ref{Figure 3}(i). In this case (S-D separation of 1.5 cm), some of light interacted with the cerebral blood flow, as the previous research from Selb \textit{et al.}\cite{Selb14Neurophot} showed that the blood flow change could be seen at an S-D separation larger than 1 cm. The increase of rCBF values in Phase 3 at the S-D separation of 7.5 mm (case (iii)) was not as clear as that in case (i), as shown by the red line in Fig.\! \ref{Figure 3}(i). In this case (S-D separation of 0.75 cm), most of light interacted with the forehead skin rather than the brain, hence the breath holding task did not have the same significant impacts on the signal. The normal breathing measurements at the S-D separation of 1.5 cm (case (iii)) as a reference did not have significant change of rCBF, as shown by the yellow line Fig.\! \ref{Figure 3}(i).

\section{DISCUSSION}

Here we demonstrated the concept of ISVS, characterized the performance of ISVS and implemented it in human cerebral blood monitoring. The interferometric detection and camera parallel measurement of multiple speckles allowed us to sample the weak signal light with high fidelity and a high frame rate. By using the sensitive ISVS system, we monitored the pulsatile blood flow in brain as well as the cerebral blood flow change during a breath holding task. The photon rate in our demonstration was $\sim\!$ 100 fold less than the photon rate in conventional DCS measurements.

\begin{table*}
\begin{ruledtabular}
\caption{\label{Table 1}Overview of the mainstream non-invasive CBF imaging techniques. The last column is added to place ISVS in perspective.}
\begin{tabular}{cccccccccc}
 &O\textsuperscript{15}PET&SPECT&XeCT&CT-P&DCS-MRI&ASL-MRI&Doppler Ultrasound&DCS&ISVS\\ \hline
 Bedside&No&Sometimes&No&No&No&No&Yes&Yes&Yes\\
 Contrast Agent&Yes&Yes&Yes&Yes&Yes&No&No&No&No\\
 Radiation&Yes&Yes&Yes&Yes&No&No&No&No&No\\
 Acq. Time&5--9 min&10--15 min&10 min&40 sec&1 min&5--10 min&10--20 min&0.5--6 sec&1--10 ms\\
 Parameters&CBF&CBF&CBF&MTT&MTT&CBF&BFV&CBF&CBF\\
 Quantitative&Yes&Sometimes&Yes&Yes&N/A&Yes\footnote{indicates measurement limitations when CBF $\!<\!10$ ml/min/100 g.}&N/A&Relative\footnote{ indicates that absolute measurements may be possible. The acronym MTT stands for mean-transit time, BFV for BF velocity, DSC-MRI for dynamic susceptibility weighed MRI, CT-P for perfusion CT.}&Relative\\
 Spatial Resolution&$\sim\!$ 5 mm&$\sim\!$ 5 mm&$\sim\!$ 5 mm&$\sim\!$ 1.5 mm&$\sim\!$ 2 mm&$\sim\!$ 5 mm&N/A&$\sim\!$ 10 mm&$\sim\!$ 10 mm\\
 Intrascan time&10 min&10 min&20 min&10 min&25 min&0 min&0 min&0 min&0 min\\
 Instrument Cost&High&High&Moderate&Moderate&High&High&Low&Low&Low
\end{tabular}
\end{ruledtabular}
\end{table*}

From a more general perspective, diffusing correlation spectroscopy-based methods can in general be categorized as temporal sampling (DCS and interferometric DCS\cite{Zhou18Optica}) or spatial ensemble (SVS and ISVS) methods. Both of them use corresponding mathematical models to describe the light-tissue interaction and speckle statistics to infer the temporal dynamics of signals of interest, such as cerebral blood flow. temporal sampling methods (e.g., DCS, interferometric DCS) directly sample one or several dynamical speckles temporally using high speed detectors, while spatial ensemble methods, including ISVS, sample speckles spatially using detector arrays. Since the SNR of the decorrelation time measurement is tied to the number of signal photons, getting a higher SNR usually requires more signal photons. The more signal photons in temporal sampling methods inevitably leads to longer measurement time, while in spatial ensemble methods that leads to more camera pixels. Therefore, in spatial ensemble methods, scaling up of the number of camera pixels will scale up the SNR of signal data points without increasing the measurement time, given high enough data throughputs. 

Another property of ISVS is that it does analysis on complex fields directly. As shown in Eq. (\ref{Eq3}), the off-axis holography scheme allows us to reconstruct the complex field and incorporate the field correlation function $g_1(t)$ directly into the visibility factor. In conventional methods such as temporal sampling DCS or spatial ensemble SVS, it is usually necessary to measure the intensity and use the Siegert relation\cite{Siegert1943} to get the information of the field correlation function $g_1(t)$. There are several conditions required for the Siegert relation to be valid, such as the Gaussian statistics and ergodicity requirements for the speckle field\cite{Borycki16Optica}. When light interacts with static parts in tissue and diffuses to the detection plane, such as surface reflection or shallow skin diffusion, there will be a static speckle field adding on the dynamic speckle field. This static speckle field breaks the Gaussian statistics and ergodicity requirements, and the Siegert relation in this case does not hold. ISVS circumvents the Siegert relation and is able to directly retrieve the information of the field correlation function $g_1(t)$.

The goal for future development of ISVS is to improve the stability of the light collection part in the system. As we use a large core multimode fiber to collect the signal light, it is highly sensitive to environmental perturbation, such as vibration. The perturbation can cause significant changes in mode mixing in the fiber and results in a decorrelating speckle pattern at the output of the fiber even if the input optical field of the fiber is static. If the speckle decorrelation time caused by environmental perturbation is comparable to the signal light decorrelation time, the contribution from environments can overwhelm the signal itself. One viable solution to this problem is to use a fiber bundle to replace the large core multimode fiber. The mode mixing between the individual fibers in the bundle is less than that in the large core multimode fiber, thus it is less sensitive to environmental perturbation.

Before concluding the discussion section, we would like to speculate on the potential role of ISVS within the currently used clinical methods of accessing the information of CBF. Table \ref{Table 1}, which is adapted from Refs. \citenum{Durduran14Neuroimage} and \citenum{, Wintermark05Stroke}, outlines the features of the various non-invasive CBF measurement modalities, including ISVS. Since ISVS shares the similar light illumination and collection architectures with DCS, it inherits multiple advantages brought from DCS, such as bedside availability, endogenous contrast, radiation free, short data acquisition time and low cost. Due to the multi-channel property brought by the camera, the data acquisition time for each measurement is significantly shortened compared to DCS. The shortened data acquisition time $T_\mathrm{acq}$ could help improve the sampling bandwidth up to $1/T_\mathrm{acq}$, which can be up to hundreds or even thousands of Hz. In some cognitive tasks, such as cerebral blood flow signals during walking\cite{Gatouillat15BBF}, high sampling speeds are required to record the quickly-changing signals for data analysis.

\begin{acknowledgments}
We would like to thank Prof. John O’Doherty, Prof. Adoph Ralphs, Dr. Yan Liu, Dr. Haowen Ruan and Dr. Haojiang Zhou for their helpful discussions. This research was supported by Kernel – Brain Research and Technologies – Grant No. FS 13520230.
\end{acknowledgments}

The data that support the findings of this study are available from the corresponding author upon reasonable request.

\nocite{*}
\bibliography{ISVS}

\end{document}


\preprint{AIP/123-QED}

\title{Interferometric speckle visibility spectroscopy (ISVS) for human cerebral blood flow monitoring }

\author{J. Xu} \thanks{These authors contributed equally to this work.}
\author{A. K. Jahromi} \thanks{These authors contributed equally to this work.}
\author{J. Brake} \thanks{These authors contributed equally to this work.}
\affiliation{Department of Electrical Engineering, California Institute of Technology}
\author{J. E. Robinson}%
\affiliation{Division of Biology and Biological Engineering, California Institute of Technology}%

\author{C. Yang}
\email{chyang@caltech.edu.}
\affiliation{Department of Electrical Engineering, California Institute of Technology}

\date{\today}

\maketitle

\section{Supplementary note}
\subsection{Optical setup}
The optical setup of the ISVS system is shown in Supplementary Fig. S2. The beam from the laser (CrystaLaser, CL671-150) is split into a reference beam and a sample beam by a polarized beam splitter. The reference beam is coupled into a single mode fiber (FB1, Thorlabs, PM460-HP) for spatial filtering. The filtered beam is collimated by a single lens (L1) and illuminates on the camera (Phantom S640). The sample beam is coupled into a multimode fiber (FB2, Thorlabs M31L02) and the output beam is collimated and illuminates on the forehead of the human subject. The diffused light from the human subject is collected by a large core multimode fiber (Thorlabs M107L02). The output light field of the large core fiber is relayed onto the camera by a 4-f system (L2 and L3). A beam splitter (BS2) combines the reference beam and sample beam. A custom designed aperture (AP) is put on the Fourier plane of the 4-f system. A polarizer (P) is put in the sample arm to filter out the cross-polarization portion of the diffused light. To record the scattered light from the sample using conventional DCS methods, a beam splitter (BS1) is added in front of BS2 and an SPCM (PerkinElmer, SPCM-AQRH-14) records the scattered light intensity.

\subsection{System implementation on rat experiments}
In the dorsal skin flap blood flow monitoring experiments, isoflurane (1-5\%) administered in an induction box followed by maintenance on a nose cone was used to induce anesthesia on a regular laboratory rat. The dorsal skin flap of the rat was shaved and clipped on a glass slide. The rat was put on a 3-D translational stage and the illumination beam illuminated the dorsal skin flap. A 4-f system (L2 and L3, Supplementary Fig. S2B) in the optical setup imaged the skin that diffused light. 

In the dorsal skin flap blood flow monitoring experiments, ketamine 80-100 mg/kg and xylazine 8-10 mg/kg given via the intraperitoneal route was used to anesthetize a regular laboratory rat. The skin on the head and the scalp on top of the skull were surgically removed. The rat was put on a 3-D translational stage and the bregma and lambda areas were identified and illuminated by a collimated beam. A 4-f system (L2 and L3, Supplementary Fig. S2B) in the optical setup imaged the part of skull that diffused light. The distance between the illumination spot and the imaging field of view was set about 1 cm. 

\subsection{Fourier plane aperture design}
To maximize the bandwidth of the signal in the spatial frequency domain, we specifically designed the shape of the aperture on the Fourier plane of the light collection 4-f system (L2 and L3, Supplementary Fig. S2A). This rectangular shape (shown in Supplementary Fig. S4A) is different from conventional circular aperture shapes (shown in Supplementary Fig. S4B), since here we cared primarily about collecting the maximum number of speckles rather than isotropic resolution in conventional imaging. The lateral size of the aperture was designed to avoid aliasing when performing off-axis holography.  

\subsection{The mathematical derivation of visibility factor}
In this section, we derive the second moment of the interference term $S(\mathbf{r})$ in Eq. 4 in the main text. Previous work from \cite{Hussain18JB} has shown similar theoretical derivation as shown below, where they analyzed the interferometric detection for dynamic speckles in ultrasound modulated optical tomography.

 \begin{eqnarray}\label{Eqs1}
&& \langle|S(\mathbf{r})|^2\rangle = \langle S(\mathbf{r})\times S^{*}(\mathbf{r})\rangle \nonumber\\&&= \frac{1}{T} \bigg\langle\int_0^T |E_\mathrm{R}| |E_\mathrm{S}(\mathbf{r},t_1)| e^{i\phi_\mathrm{S}(\mathbf{r},t_1)} dt_1  \times \int_0^T |E_\mathrm{R}| |E_\mathrm{S}(\mathbf{r},t_1)| e^{-i\phi_\mathrm{S}(\mathbf{r},t_1)}\bigg\rangle dt_1 \nonumber\\&&= \frac{I_R}{T^2} \int_0^T \int_0^T dt_1 dt_2 \big\langle|E_\mathrm{S}(\mathbf{r},t_1)|e^{i\phi_\mathrm{S}(\mathbf{r},t_1)} |E_\mathrm{S}(\mathbf{r},t_2)|e^{-i\phi_\mathrm{S}(\mathbf{r},t_2)}\big\rangle \nonumber\\&&=\frac{I_R}{T^2} \int_0^T \int_0^T dt_1 dt_2 \big\langle E_\mathrm{S}(\mathbf{r},t_1) E^*_\mathrm{S}(\mathbf{r},t_2)\rangle \nonumber\\&&= \frac{I_R \bar{I_S}}{T^2} \int_0^T \int_0^T dt_1 dt_2 ~g_1(t_1-t_2)
\end{eqnarray} 
Here $g_1(t)$ is the field decorrelation function. Change the integration variable by $ \left\{  \begin{array}{ll}
                  t = t_1\\
                  \tau = t_1 - t_2
                \end{array}
              \right.$, 
the integration in Eq. S1 can be written as 
 \begin{eqnarray}\label{Eqs2}
&& \int_0^T \int_0^T dt_1 dt_2 g_1(t_1-t_2) = \int_0^T \int_{-t}^t d\tau g_1(\tau) \nonumber \\&&= \int_0^T \int_{\tau}^T dt g_1(\tau) + \int_{-T}^0 \int_{-\tau}^T dt g_1(\tau) \nonumber \\&&= \int_0^T 2(1-\frac{\tau}{T})g_1(\tau) d\tau = \int_0^T 2(1-\frac{t}{T})g_1(t)
\end{eqnarray} 
given $g_1(t)$ is symmetric. Therefore, Eq. S2 can be reduced to 
\begin{eqnarray}\label{Eqs3}
    \langle|S(\mathbf{r})|^2\rangle = \frac{I_R \bar{I_S}}{T^2} \int_0^T 2(1-\frac{t}{T})g_1(t)
\end{eqnarray} 
The visibility factor is defined as $F = \frac{S(\mathbf{r})^2}{I_R \bar{I_S}} = \frac{1}{T} \int_0^T 2(1-\frac{t}{T})g_1(t)$  (shown in Eq. 5 in the main text).
If $g_1(t)$ has a form of $g_1(t) = e^{-t/\tau}$  where $\tau$ is the speckle decorrelation time, after substituting in Eq. S3, we can get 
\begin{eqnarray}\label{Eqs4}
    F = \frac{S(\mathbf{r})^2}{I_R \bar{I_S}} = \frac{2\tau}{T}[1+\frac{\tau}{T}(e^{-\tau/T}-1)]
\end{eqnarray} 
As $\tau\ll T$, $\tau/T\ll 1$ and the visibility factor $F$ converges to $2\tau/T$ . 

In real cases, the visibility factor calculated from Eq. S4 has a non-zero offset. Even when no signal light photons hit the camera, there is still spatial fluctuations in the cropped side lobes due to camera noise. Taking square and summing up all the pixels give a non-zero offset that is related to camera noise. Therefore, in the main text Fig. 2B, we use a camera noise corrected model to describe the visibility factor. The red curve in Fig. 2B has the expression of the corrected visibility factor $F$ as
\begin{eqnarray}\label{Eqs4}
    F = (1-\beta) + \beta \frac{2\tau}{T}[1+\frac{\tau}{T}(e^{-\tau/T}-1)]
\end{eqnarray} 
where $\beta$ equals to 0.9.

\bibliography{ISVS}

\newpage
\section{Supplementary figures}

\begin{figure}[h!]
\centering\includegraphics[width=0.8\textwidth]{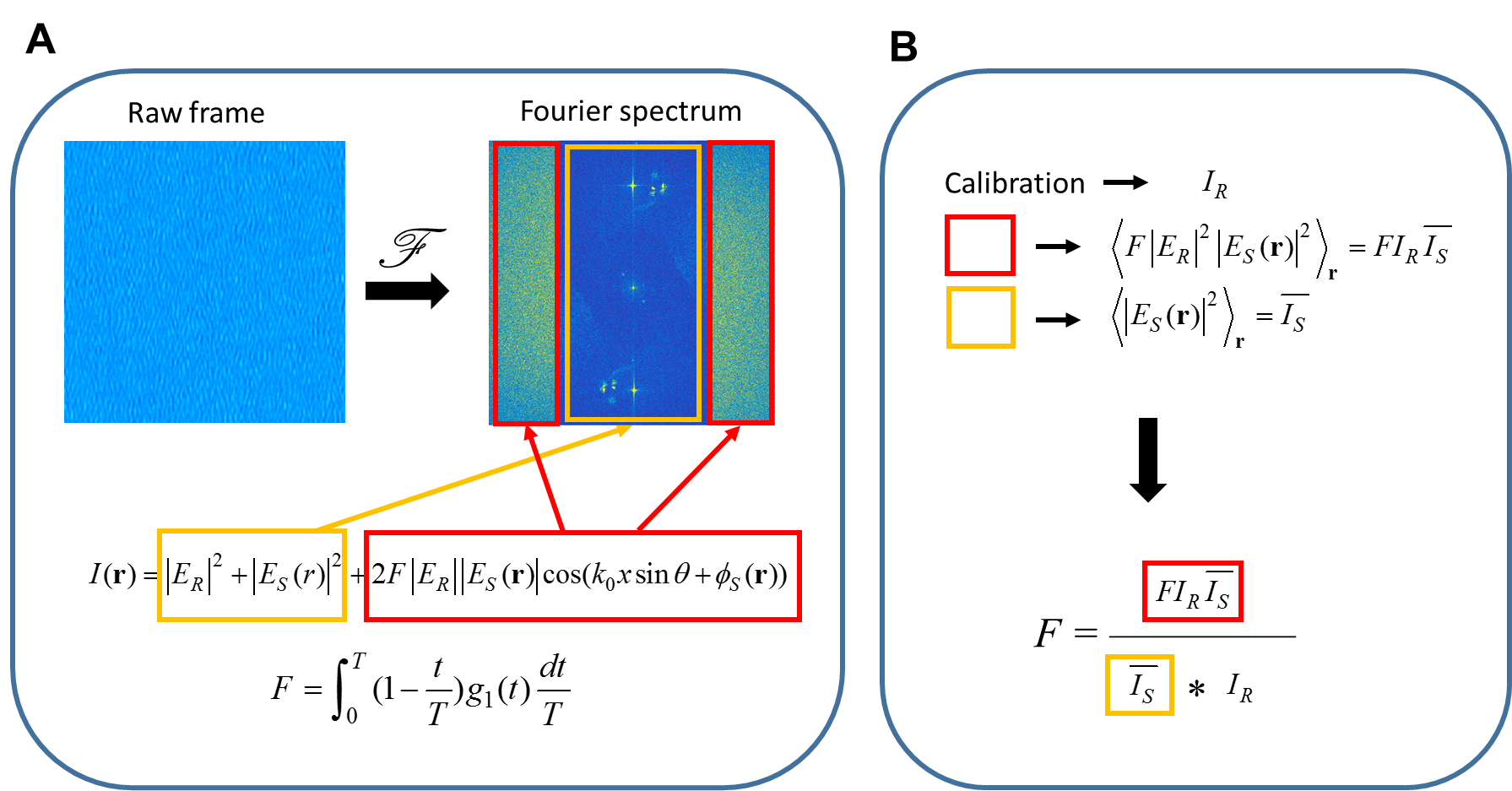}
\caption{\label{figs1} Flowchart of data processing. (A) The raw frame is Fourier transformed and three lobes contain the information of the reference beam and sample beam. The yellow enclosed circle and rectangle contain the intensity information of the reference beam and sample beam, and the red enclosed circle and rectangle contain the information of the visibility factor and the complex fields. (B) The reference beam calibration provides the mean value of the reference beam intensity, the red enclosed rectangle in (A) provides the mean value of the energy in one of the interference lobe, and the yellow enclosed rectangle in (A) provides the mean value of the sample beam intensity after subtracting the reference beam mean intensity. $<\cdot>$  denotes the ensemble average over space. }
\end{figure}

\begin{figure}[p]
\centering\includegraphics[width=0.6\textwidth]{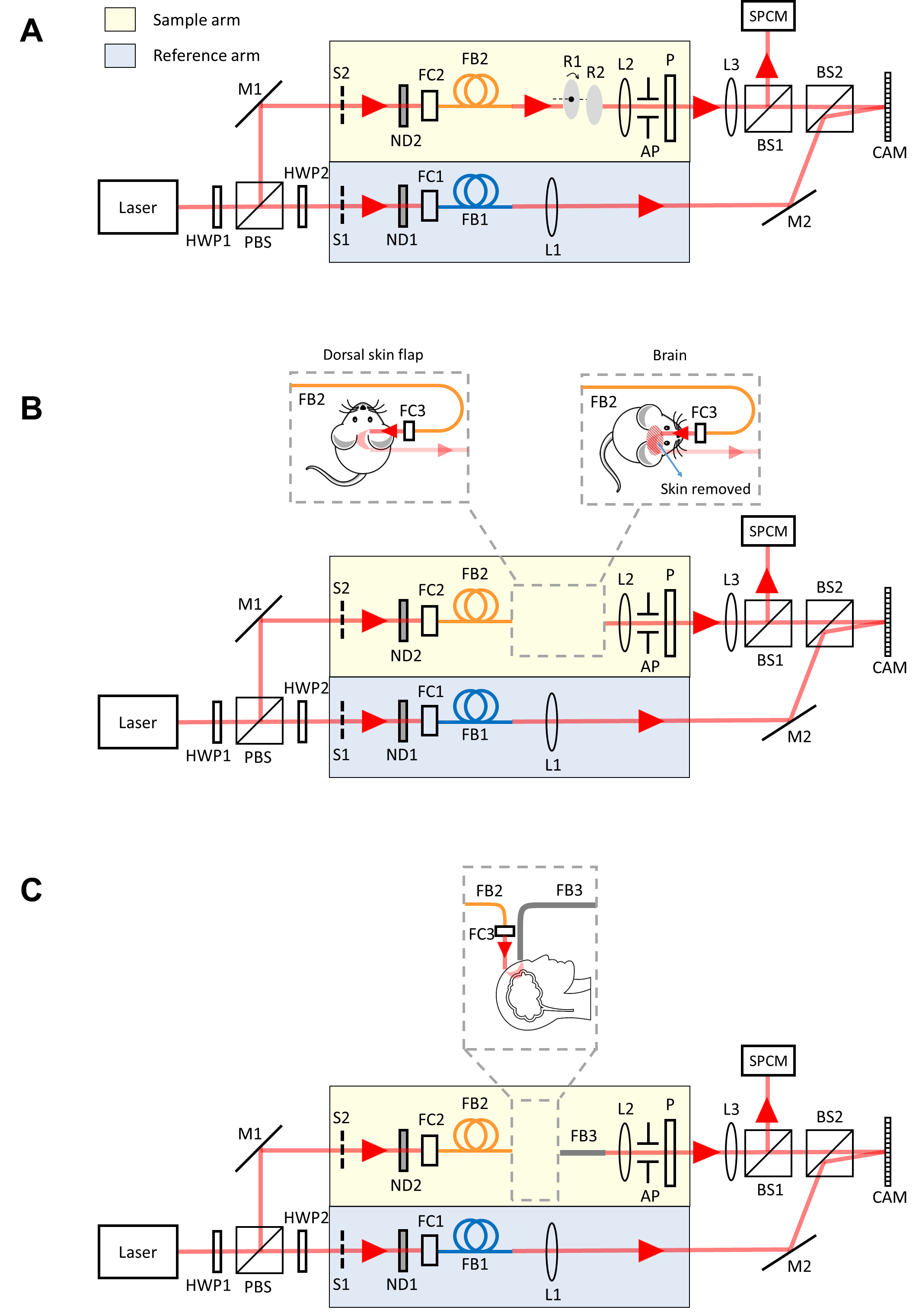}
\caption{\label{figs2} Schematic of the optical setup. (A) Optical setup for diffuser experiments. (B) Optical setup for animal experiments. (C) Optical setup for human experiments. AP, aperture; BS, beam splitter; CAM, camera; FB, fiber; FC, fiber coupler; HWP, half-wave plate; L, lens; M, mirror; ND, neutral-density filter; P, polarizer; SPCM, single photon counting module.}
\end{figure}

\begin{figure}[p]
\centering\includegraphics[width=0.8\textwidth]{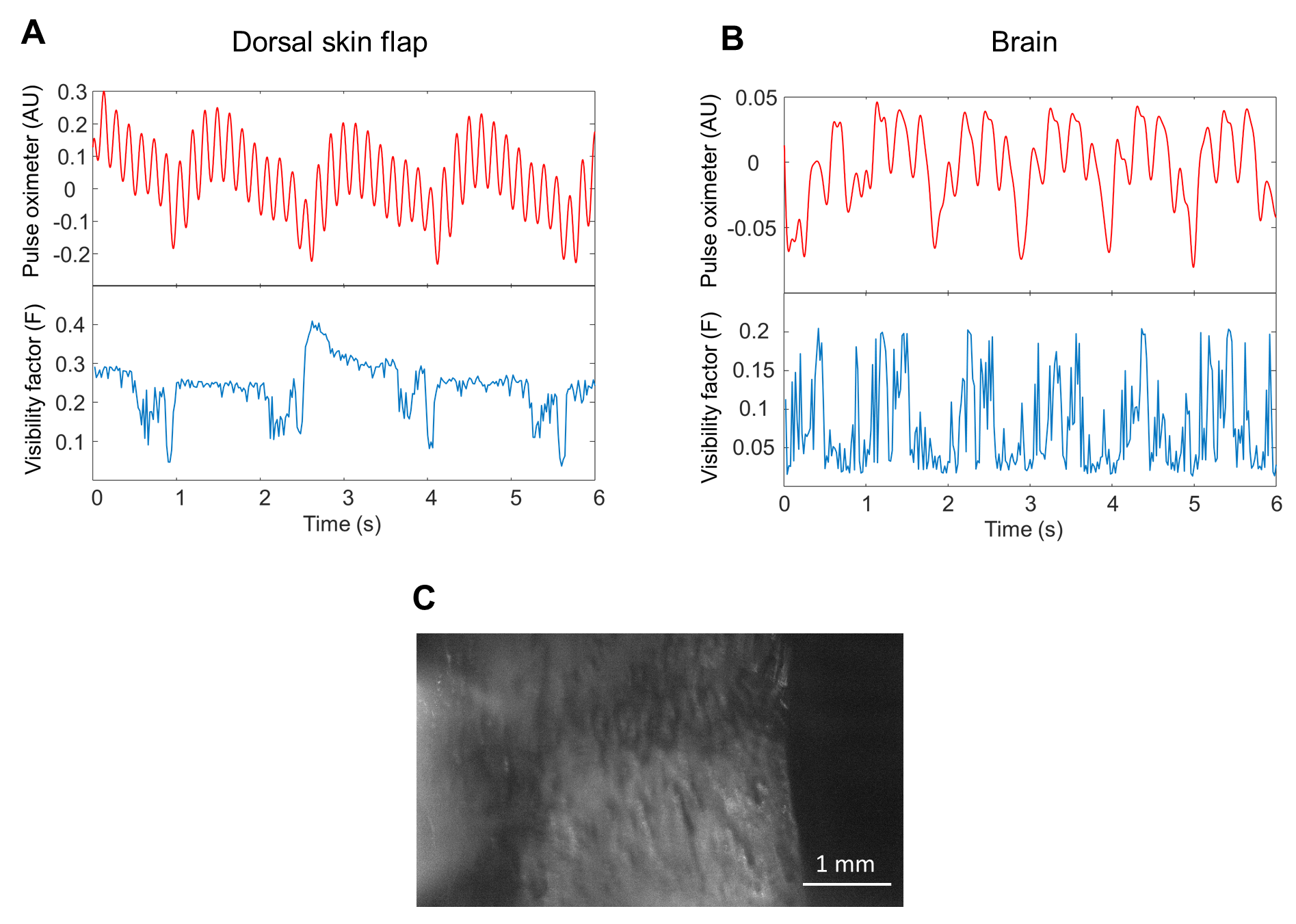}
\caption{\label{figs3} Results of rat experiments. (A) The results clearly show the breathing signal at ~1.5 Hz from both the pulse oximeter and ISVS visibility factor signals (four big dips). Pulsatile signals at ~3.5 Hz are shown in the pulse oximeter but not clearly shown in ISVS from the dorsal skin flap. This might be due to less arterial vessels in dorsal skin flap. (B) The results show the breathing signal at ~1.0 Hz from both the pulse oximeter and ISVS visibility factor signals. Pulsatile blood flow signals in ISVS couple with the breathing signals in brain signal measurements. (C) The wide field image of the brain bregma area of the rat using white LED illumination. The pulse oximeter in both measurements samples at 1 kHz. The camera frame rate in the ISVS system in both measurements is set at 50 Hz and the exposure time is set as 16 ms.}
\end{figure}

\begin{figure}[hbt!]
\centering\includegraphics[width=0.6\textwidth]{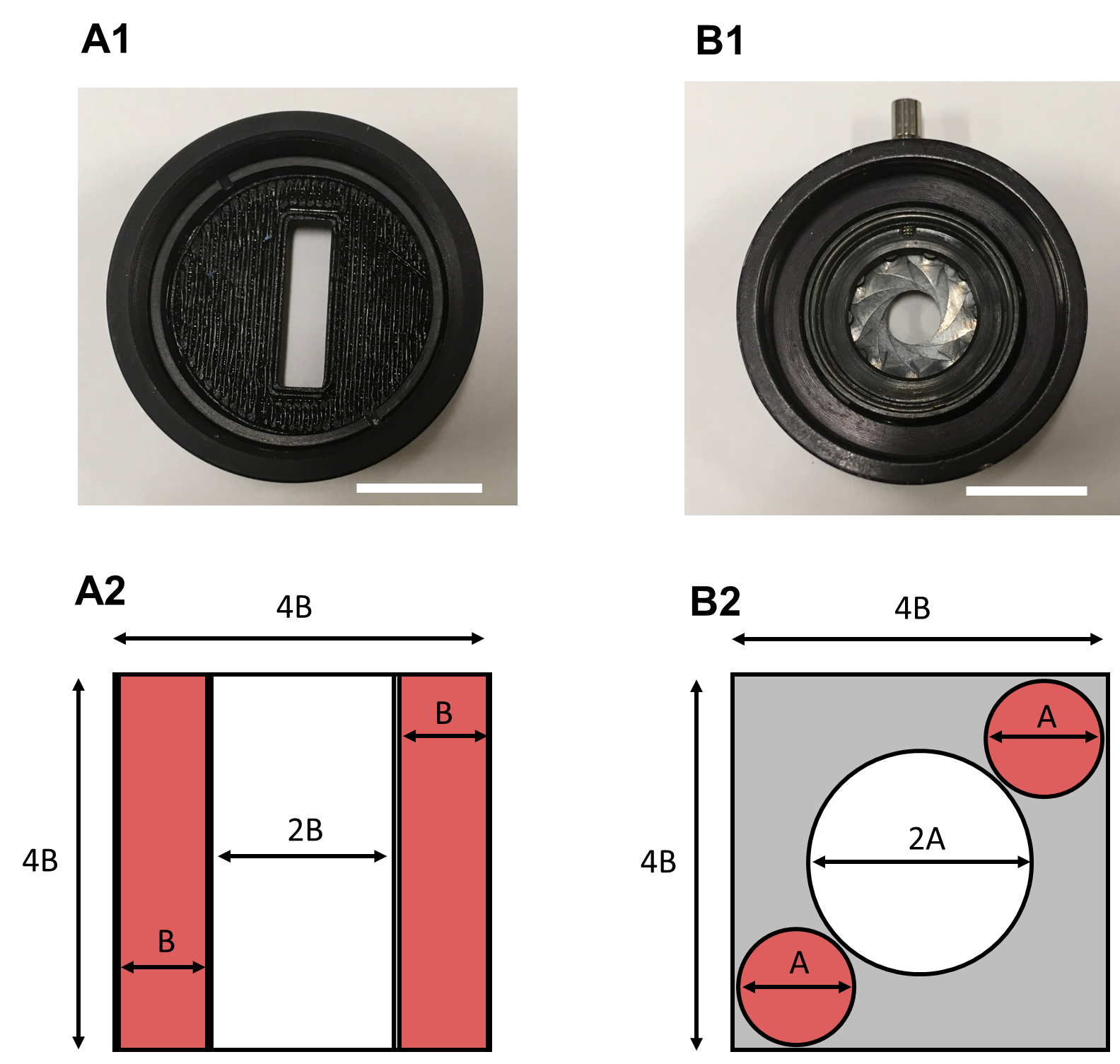}
\caption{\label{figs4} Fourier plane aperture design. (A1) Rectangular aperture. (A2) The Fourier spectrum of an off-axis hologram with a rectangular aperture. (B1) Circular aperture. (B2) The Fourier spectrum of an off-axis hologram with a circular aperture. Mathematically, it can be shown that $A = \frac{4\sqrt{2}}{3+\sqrt{2}}B \approx 1.28B $. Therefore, the circular aperture uses $\frac{2\pi A^2/4}{(4B)^2} \approx 16\%$  (red circles in (B2) of the Fourier space while the rectangular aperture uses 50\% (red rectangles in (A2)) of the Fourier space. The larger area in Fourier space allows higher light collection efficiency. Scale bar, 1 cm.}
\end{figure}
